\documentclass[conference]{IEEEtran}
\IEEEoverridecommandlockouts
\usepackage{cite}
\usepackage{amsmath,amssymb,amsfonts}
\usepackage{algorithm}
\usepackage{algorithmic}
\usepackage{graphicx}
\usepackage{textcomp}
\usepackage{xcolor}
\usepackage{multirow}
\usepackage{subcaption}
\usepackage{fancyhdr}
\usepackage{booktabs}
\usepackage {makecell} 

\newcommand{\tco}{\textcolor{black}}
\def\BibTeX{{\rm B\kern-.05em{\sc i\kern-.025em b}\kern-.08em
    T\kern-.1667em\lower.7ex\hbox{E}\kern-.125emX}}
\begin{document}

\title{Diagnosis-oriented Medical Image Compression with Efficient Transfer Learning}


\author{Guangqi Xie\textsuperscript{*}, Xin Li\textsuperscript{*}, Xiaohan Pan, Zhibo Chen$^\dagger$ \\
\textit{Unviersity of Science and Technology of China, Hefei, China} \\
\small \{jszjxgq, lixin666, pxh123\}@mail.ustc.edu.cn,  chenzhibo@ustc.edu.cn}
\newcommand{\ieno}{\textit{i.e.}}
\newcommand{\egno}{\textit{e.g.}}

\maketitle
\thispagestyle{fancy}
\renewcommand{\thefootnote}{\fnsymbol{footnote}}
\footnotetext[1]{Equal Contribution}
\footnotetext[2]{Zhibo Chen is corresponding author}
\fancyhead{}
\renewcommand\headrulewidth{0pt}

\begin{abstract}
Remote medical diagnosis has emerged as a critical and indispensable technique in practical medical systems, where medical data are required to be efficiently compressed and transmitted for diagnosis by either professional doctors or intelligent diagnosis devices. In this process, 
a large amount of redundant content irrelevant to the diagnosis is subjected to high-fidelity coding, leading to unnecessary transmission costs. To mitigate this, we propose diagnosis-oriented medical image compression, a special semantic compression task designed for medical scenarios, targeting to reduce the compression cost without compromising the diagnosis accuracy. However, collecting sufficient medical data to optimize such a compression system is significantly expensive and challenging due to privacy issues and the lack of professional annotation. In this study, we propose DMIC, the first efficient transfer learning-based codec, for diagnosis-oriented medical image compression, which can be effectively optimized with only few-shot annotated medical examples, 
by reusing the knowledge in the existing reinforcement learning-based task-driven semantic coding framework, \ieno, HRLVSC~\cite{xie2022hierarchical}. Concretely, we focus on tuning only the partial parameters of the policy network for bit allocation within HRLVSC, which enables it to adapt to the medical images. In this work, we validate our DMIC with the typical medical task, Coronary Artery Segmentation. Extensive experiments have demonstrated that our DMIC can achieve 47.594\%BD-Rate savings compared to the HEVC anchor, by tuning only the A2C module (2.7\% parameters) of the policy network with only 1 medical sample. 

\end{abstract}


\section{Introduction}

Nowadays, remote medical diagnosis/telemedicine\cite{kaur2015roi,juliet2015projection,nasifoglu2019medical} is gaining popularity across various application scenarios, including rural healthcare\cite{bulsara2011low}, emergency consultations, international collaborations, post-operative follow-up, and home healthcare monitoring, etc. In these scenarios, the highly efficient transmission of remote medical data is of great importance and indispensable\cite{sheeja2020soft,dutta2014medical}. However, for specific medical diagnostic tasks, there is a significant amount of non-essential redundant information present in the raw medical data that does not affect the diagnostic outcomes\cite{kaur2015roi,hu2008multi}. Consequently, the high-fidelity compression of these diagnosis-irrelevant content usually causes unnecessary occupation of substantial bandwidth, imposing significant pressures on data transmission and storage\cite{mitra1998wavelet}.

To alleviate the above challenges, we propose the diagnosis-oriented medical image compression task, where the focus is to reduce the compression cost while preserving the diagnosis accuracy. Notably, it can be viewed as a special semantic compression task in medical scenarios, \ieno, the diagnosis-related semantics, which motivates us to seek the solution from the perspective of semantic coding. Existing works on semantic coding can be roughly divided into two categories based on the types of the basic codec. The first category~\cite{DeepSIC,DSSLIC,EDMS,erdemir2023generative,huang2021deep} primarily focuses on semantic coding within the framework of layered end-to-end compression. However, it is still far away from the wide application for end-to-end compression in practical scenarios. Consequently, the second line~\cite{lu2022preprocessing,shi2020reinforced,li2021task,xie2022hierarchical} aims to achieve the semantic coding based on the traditional hybrid compression framework, like HEVC~\cite{sullivan2012overview}, VVC~\cite{bross2021overview} with pre-processing~\cite{lu2022preprocessing} or reinforcement learning~\cite{shi2020reinforced,li2021task,xie2022hierarchical}, etc. In particular, the most advanced semantic coding technique of the second line is HRLVSC~\cite{xie2022hierarchical}, which introduces a groundbreaking solution for video semantic coding based on hierarchical reinforcement learning. Specifically, it leverages reinforcement learning to address the challenge of non-differentiability in video semantic coding, and then introduces hierarchical reinforcement learning to decouple the complex mode decision space into frame- and CTU-level.
Despite that, these algorithms, designed specifically for natural image coding, cannot be directly applied to medical image coding due to the huge distribution gap.

\begin{figure*}
\centering 
\includegraphics[width=1.0\textwidth]{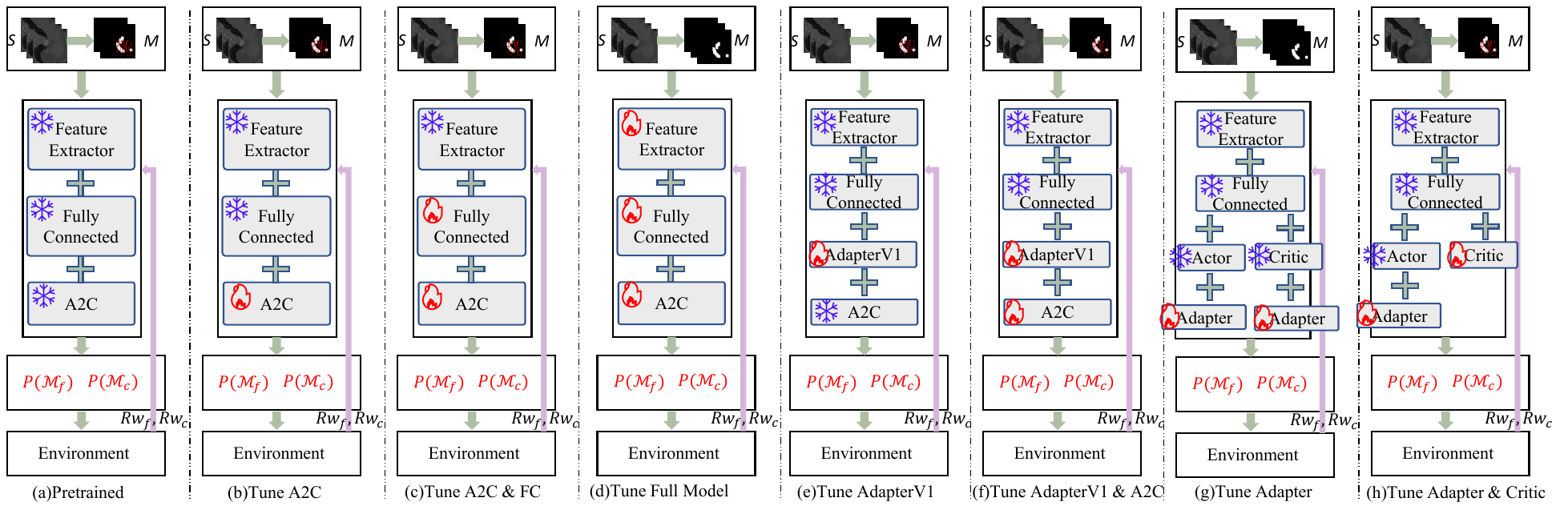}

\caption{Illustration of our proposed DMIC. From left to right: (a) Pre-trained model. (b) Tuning only A2C module, which is used in this work. (c) Tuning the A2C and the fully connected layer. (d) Tuning the full model. (e) Adding an adapter called AdapterV1 for the fully connected layer and then tuning it. (f) Tuning both the AdapterV1 and A2C. (g) Adding an Adapter after both the actor and critic, and tuning them. (h) Adding an Adapter after the actor, and tuning both the critic and the Adapter.}
\label{fig:Network}
\vspace{-5mm}
\end{figure*}

An intuitive solution for diagnosis-oriented medical image compression is to retrain the above semantic codecs with medical images. However, it is inaccessible to collect enough medical data for optimization due to privacy issues and the lack of professional annotation\cite{bian2021domain,lu2022rtn}. To overcome this, we propose the first efficient transfer learning-based semantic codec, intending to adapt the prior knowledge from the pre-trained HRLVSC codec with abundant natural images to the diagnosis-oriented medical image compression in specific medical scenarios. It is noteworthy that the only trainable key modules of HRLVSC are the frame-level and CTU-level RL agents. And thus, we aim to seek an efficient and effective tuning strategy for HRLVSC by tuning the partial parameters of the RL agents with few-shot medical data, thereby achieving the efficient transfer learning-based semantic codec for \textbf{D}iagnosis-oriented \textbf{M}edical \textbf{I}mage \textbf{C}ompression, named DMIC.  
To be more detailed, we have explored various tuning strategies aimed at striking a balance between the number of tuning parameters, training cost, and model performance. 
Through careful experimentation and analysis, we aim to provide practical insights into achieving optimal performance in diagnosis-oriented medical image compression task despite data scarcity.

The main contributions of our work can be summarized as follows:

\begin{itemize}
    \item  \tco{As the pioneering work, we propose DMIC, the first efficient transfer learning-based semantic codec for diagnosis-oriented medical image compression, intending to transfer the prior knowledge from pre-trained HRLVSC to downstream medical compression task by tuning few parameters with few-shot medical data.}

    \item \tco{We conduct an in-depth analysis of different tuning strategies and propose a simple yet effective tuning mechanism for HRLVSC. Our approach achieves comparable performance while tuning only the A2C module (2.7\% parameters) of the policy network.}
 
    \item Extensive experiments conducted under the low-delay P configuration demonstrate the superiority of our proposed solution. We outperform the standard software HM16.24 by achieving a BD-rate saving of 47.594\% for the Coronary Artery Segmentation task.
\end{itemize}

The rest of the paper is organized as follows. In sec.~\ref{sec: method}, we clarify our DMIC in detail. Sec.~\ref{sec: experiments} describes our experimental setting and validates the effectiveness of our proposed DMIC by comparing it with the state-of-the-art codecs and a series of ablation studies. Finally, we conclude this paper in Section~\ref{sec: conclusion}.

\section{Diagnosis-oriented medical
image compression}
\label{sec: method}

In this section, we first clarify the framework of our DMIC based on HRLVSC, and then  
describe our explored efficient tuning strategies for the optimal DMIC under the limitation of few-shot medical data.

\subsection{Overall Framework}
Our DMIC is based on the framework of HRLVSC, which exploits hierarchical reinforcement learning (RL) to achieve the semantic bit allocation, thereby achieving the video semantic coding. Concretely, our DMIC contains three parts: 1) hierarchical trainable RL agents, \ieno, the frame-level agent, and CTU-level agent, that are responsible for mode decision for each frame, and CTU, respectively; 2) mask generation with the downstream task; and  3) traditional hybrid compression codec. Given a medical image from the diagnosis task, DMIC first extracts the diagnosis-related semantic masks with the diagnosis task for the medical slices. Then the medical slices and their semantic masks are jointly inputted to the frame-level and child-level RL agents, as shown in Fig.~\ref{fig:Network}(a), producing the frame-level and CTU-level mode decision, \ieno, action decision.  For frame-level mode decision, we exploit the default QP structure in HM16.24 Low Delay P GOP 8, which is in the form of $QP_{I}$ followed by a loop of $\{\Delta QP_{P}, \Delta QP_{P}-1, \Delta QP_{P}, \Delta QP_{P}-1, \Delta QP_{P}, \Delta QP_{P}-1, \Delta QP_{P}, QP_{I}+2\}$. And the action space is set for two quantization parameters, $QP_{I}$ and $\Delta QP_P$. The latter is decided based on the former, setting as $\Delta QP_{P}=6, 7, 8$  for $QP_{I}\in [14, 21], [22, 29], [30,37]$ separately. For CTU level action space, the offset $\Delta QP$ of foreground CTU is set from $\{-2,0,+2\}$ and the $\Delta QP$ of background CTU is set from $\{+2, +4, +6, +8\}$. After obtaining the optimal actions of two RL agents, we can control the bit allocation process for the traditional codec (\ieno, the environment in RL) with the above coding mode. The semantic distortion is measured with the difference of the diagnosis accuracy between the compressed medical slice and the original medical slice. In this way, we can achieve the rate-distortion optimization (\ieno, the reward) of diagnosis-oriented image compression based with reinforcement learning.  

\subsection{Efficient Transfer Learning for DMIC}
Since the lack of annotated medical images, it is inaccessible to train our DMIC from scratch. Therefore, we propose an efficient transfer learning for our DMIC, intending to exploit the prior knowledge from the natural image/video semantic task. 
As stated in the above section, our DMIC is constructed based on HRLVSC, where the key module is the frame-level and CTU-level RL agents. Consequently, we systematically investigate which part of the RL agent is optimal for the knowledge transfer to the medical image diagnosis tasks with few-shot annotated medical images. 




For uniformity, we formulate both the frame- and CTU-level agents as Eq.\ref{equ:structure}:
\begin{equation}
\begin{aligned}
    &\mathbb{S}=FC(FE(S,M(S)))\\
    &\mathbb{M}=A2C(\mathbb{S})
\end{aligned}
\label{equ:structure}
\end{equation}
$S$ means medical image slices, $M(S)$ indicates the diagnosis-oriented masks. The network of the RL agent is composed of a feature extractor, a fully connected layer, and an advantaged actor-critic network, depicted as $FE,FC,A2C$ separately. State is depicted as $\mathbb{S}$ while $\mathbb{M}$ indicates action, \textit{i.e.}, optimal modes. To facilitate data-efficient training, we propose 7 tuning strategies as shown in Fig.\ref{fig:Network} (a)$\sim$(e). In this context, the pre-trained HRLVSC model (a) is set as our baseline, representing the performance lower bound.  Afterward, the actor-critic network $A2C$, the fully connected layer $FC$, and the feature extractor $FE$ are ablated progressively as shown in Fig.~\ref{fig:Network} (b) (c) and (d), separately. Concretely, (b) tunes only the $A2C$ module, (c) tunes both the $A2C$ module and the fully connected layer, and (d) tunes the full model. 
For another four tuning strategies, we incorporate the popular adapter-style method~\cite{li2023graphadapter,gao2023clipadapter} into the different components of the RL agent, respectively.  
Particularly, (e) adds an adapter called    ``AdapterV1" for the fully connected layer for tuning. (f) tunes both the ``AdapterV1" and the A2C network. (g) adds two adapters called ``Adapter" for both the actor and critic and then only tunes them. (h) tunes the critic and the ``Adapter" for the actor. Among the 7 different tuning strategies, tuning $A2C$ module(b) is experimentally the best solution. The reason might be that the feature extractor and fully connected layer for the state are more robust for different tasks since the introduction of the semantic mask, while the $A2C$ module is required to be task-adaptive since it directly decides the reward and action for a specific task. 

\begin{figure}[ht]
\centering
\begin{subfigure}{1.0\linewidth}
\centering
\includegraphics[width=0.9\linewidth]{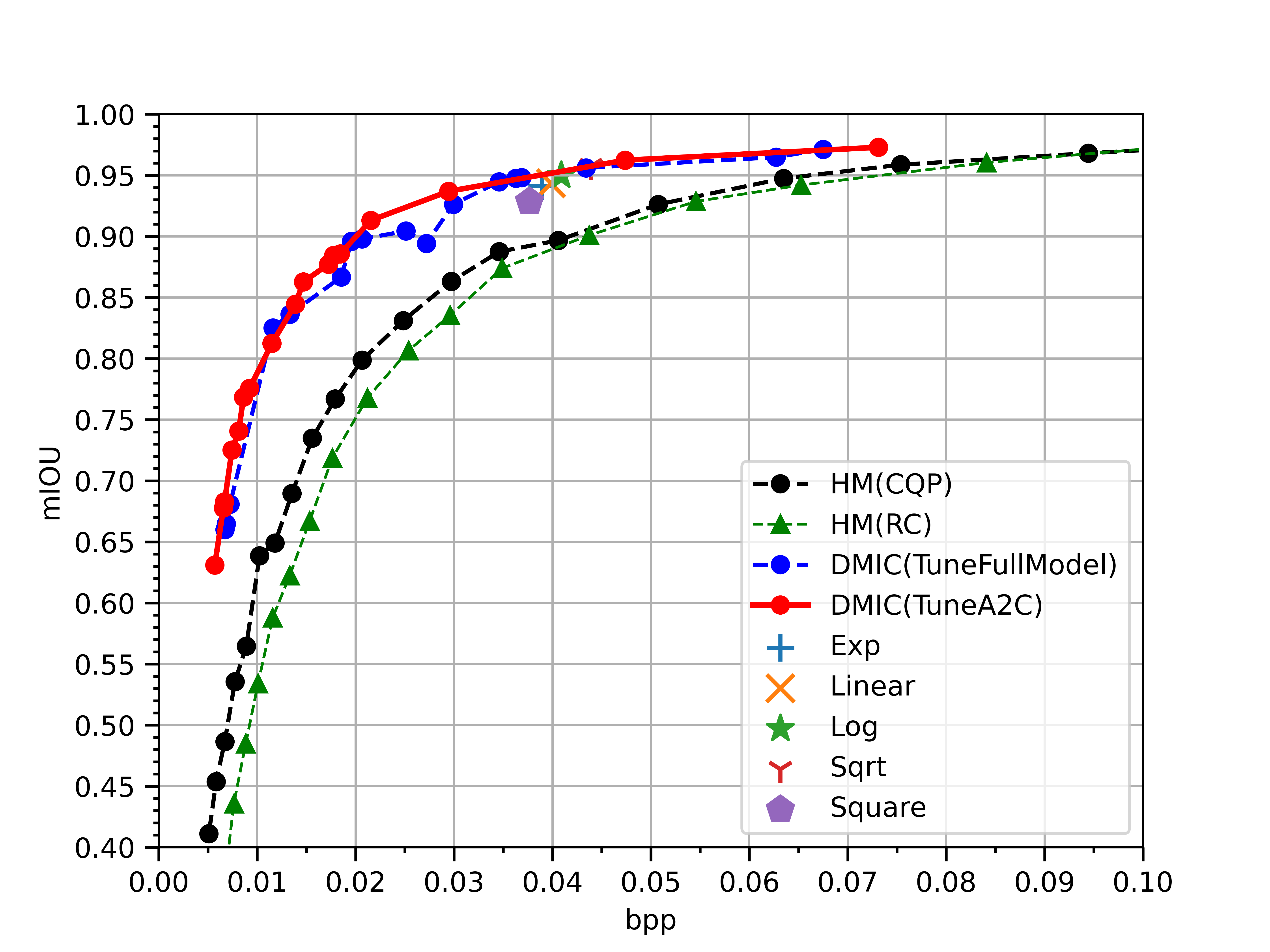}
\vspace{-3mm}
\caption{Rate-Task-Distortion Performance.}
\label{fig:RD}
\end{subfigure}
\begin{subfigure}{1.0\linewidth}
\centering
\includegraphics[width=0.9\linewidth]{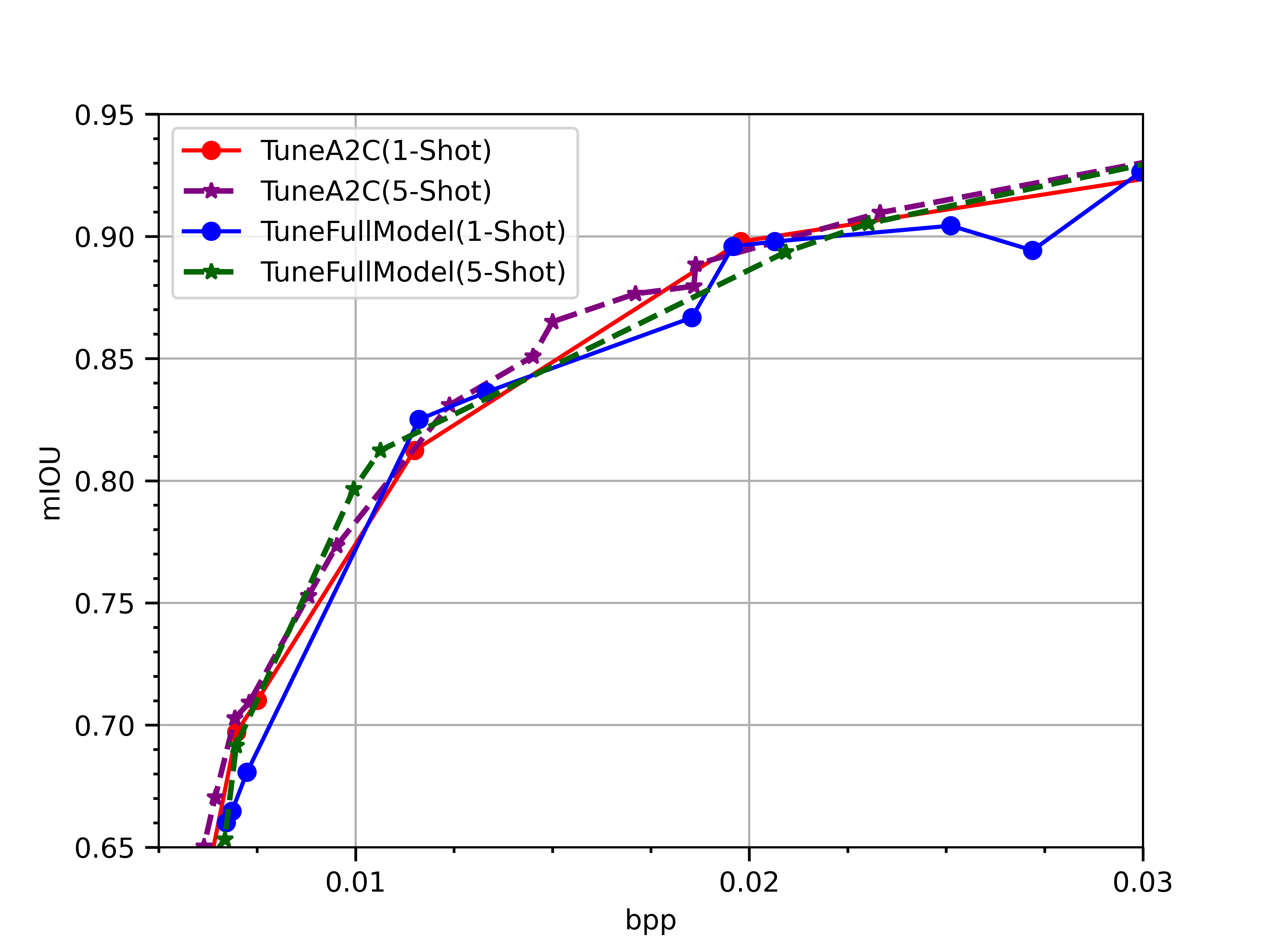}
\vspace{-3mm}
\caption{Ablation Study on Shot Number.}
 \label{fig:shot}
\end{subfigure}
\hfill
\begin{subfigure}{1.0\linewidth}
\centering
\includegraphics[width=0.89\linewidth]{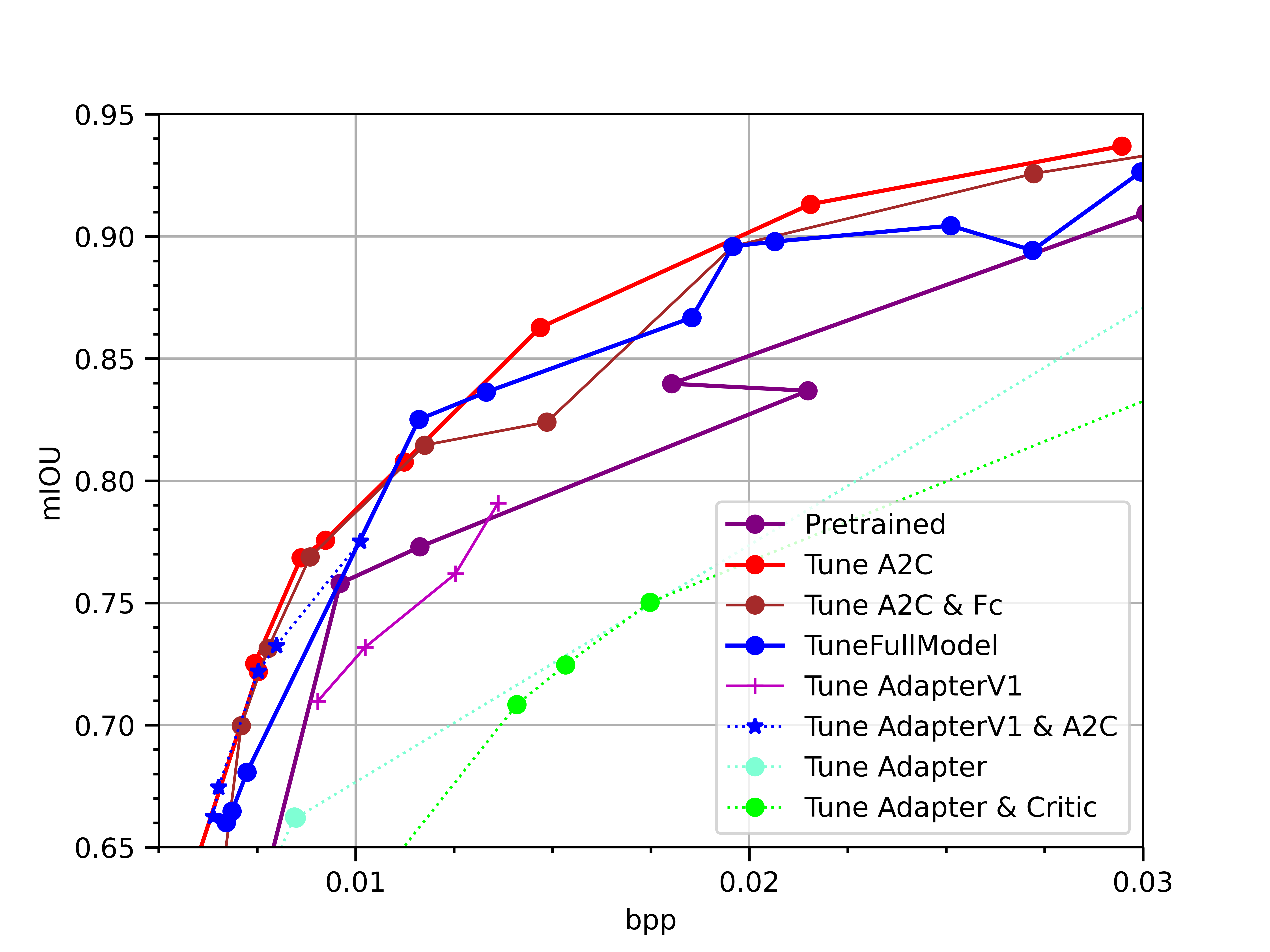}
\vspace{-3mm}
\caption{Ablation Study on Tuning Strategies under 1-shot Setting.}
 \label{fig:Tune}
\end{subfigure}
  \caption{Performance Comparison and Ablation Studies.}
  \label{fig:performance}
\end{figure}

\section{Experiments}
\label{sec: experiments}

\subsection{Dataset and Implementation Details}
To validate the effectiveness of our DMIC for Diagnosis-oriented Image Compression, we choose imageCAS\cite{zeng2022imagecas} task, \ieno, the Image Coronary Artery Segmentation task as our downstream task. 

\noindent\textbf{Dataset:} We collected the dataset of our DMIC from the imageCAS task~\cite{zeng2022imagecas}, where each slice is organized as video form, which is convenient  for the compression with the traditional codec. We compress them with the modes used in our optional action spaces, as stated in section Overall Framework. And we divide the training set, validation set, and test set according to the sample number of K:10:10. K is set from $\{1, 3, 5\}$ separately. 

\noindent\textbf{Implementation Details:}
The DMIC model is implemented with the Pytorch framework, which is trained with one NVIDIA RTX 3090 GPU for 1000 iterations. The batch size is set as the same as the shot number. The learning rate is 1e-3 for frame-level agent and 1e-4 for CTU-level agent.

\subsection{Performance Analysis}
\subsubsection{Compared with HEVC Anchor}
To verify the effectiveness of our DMIC scheme, we compare our DMIC with standard HEVC codec HM 16.24 under Low-delay P configuration.  
For HEVC, we select the QP from 12 to 42 for compression as our baseline. For our proposed DMIC, we set $\lambda$ from 0 to 100 with an increasing interval.
The result is shown in Table~\ref{tab:HRLvsAnchor}. From the table, we can observe that our proposed DMIC scheme outperforms HEVC anchor by a bit saving of 47.594\%,  which demonstrates the effectiveness of our DMIC for remote medical diagnosis. However, the rate control cannot reduce the BD-Rate, which is not proper for the semantic coding of the imageCAS task.


\begin{table}[H]
    \centering
    \vspace{-3mm}
    \caption{BD-Rate and BD-mIOU compared with Anchor HM16.24(constant QP).}
    \begin{tabular}{c|c|c|c}
        \hline
        Method & HEVC w/ rate control &  DMIC(Tune A2C) & Pretrained\\
        \hline
        BD-Rate & +15.56\% & -47.594\% & -27.41\%\\
        \hline
        BD-mIOU & -0.03\%  & +0.089\% &  +0.049\%\\
        \hline
    \end{tabular}
    \label{tab:HRLvsAnchor}
    \vspace{-3mm}
\end{table}

\subsubsection{Ablation Study}
We conduct the ablation studies from three perspectives: 1)  different shot numbers, 2) different tuning strategies, and 3) substituting our DMIC with hand-crafted methods \ieno, assigning the higher QP value for semantic-unrelated CTUs and lower QP value for semantic-related CTUs. The experimental results are as follows:

\textbf{Ablation study on shot number}. As shown in Fig.~\ref{fig:shot} and Table~\ref{tab:Shot Number}, our methods achieves consistent BD-Rate savings under different shot number, which means our DMIC can be robust to different shot numbers. The performance can be attributed to the preservation of prior knowledge within the feature extractor and the task-specific adaptations in the action and reward module (\ieno, the A2C module). Moreover, we can observe that ours can achieve the obvious gain compared with tuning the full model on different shot settings. 

To fully exploit the potential of few-shot data, we performed subsequent ablation studies in the 1-shot setting.

\begin{table}[H]
\vspace{-3mm}
    \caption{Ablation Study on Data Efficiency.}
    \centering
    \begin{tabular}{c|c|c|c}
        \hline
        \makecell[c]{Training \\ scheme} & \makecell[c]{Shot \\ Number}  & Params & BD-Rate Savings\\
        \hline
        \multirow{3}*{Tune Full Model} & 5  & 1058992 & -47.906\%\\
        \cline{2-4}
        ~ & 3   & 1058992 & -47.520\%\\
        \cline{2-4}
       ~ & 1   & 1058992 & -44.596\%\\
       \hline
        \multirow{3}*{Tune A2C} & 5 & \textbf{28784} & \textbf{-48.260\%}\\
        \cline{2-4}
       ~ & 3 & \textbf{28784} & \textbf{-47.678\%}  \\
        \cline{2-4}
         ~ & 1  & \textbf{28784} & \textbf{-47.594\%}  \\
         \hline
    \end{tabular}
    \label{tab:Shot Number}
    \vspace{-3mm}
\end{table}

        

\begin{table}[H]
\vspace{-3mm}
    \caption{Ablation Study on Different Tuning Strategies under 1-shot.}
    \centering
    \begin{tabular}{c|c|c}
        \hline
       Training Scheme  & Params & BD-Rate\\
        \hline
        
        (a)Pretrained & 0  & -27.410\%\\
        \hline
        \textbf{(b)Tune A2C}  & \textbf{28784} & \textbf{-47.594\%}\\
        \hline
       (c)Tune A2C \& FC & 946800 & -45.402\%\\
        \hline
        (d)Tune Full Model  & 1058992 & -44.596\%\\
        \hline
        (e)Tune AdapterV1 & 131584 & -15.296\%\\
        \hline
        (f)Tune AdapterV1 \& A2C  & 160368 & -45.058\%\\
        \hline
        (g)Tune Adapter & 10214 & -19.286\%\\
        \hline
        (h)Tune Adapter \& Critic  & 10724 & -15.098\%\\
        \hline
    \end{tabular}
    \label{tab:Tune}
    \vspace{-3mm}
\end{table}

\begin{table}[H]
\vspace{-3mm}
    \caption{Time complexity.}
    \centering
    \begin{tabular}{c|c|c|c}
        \hline
        Run time & \multicolumn{2}{|c|}{Encoder} & Decoder \\
        \hline
        \multirow{2}*{Proposed} & HRL agent & 0.92s & \multirow{2}*{0.203s}\\
        \cline{2-3}
        ~&coding& 124.67s &~\\
        \hline
        HEVC w/o Rate Control&\multicolumn{2}{|c|}{137.02s}&0.203s\\
        \hline
    \end{tabular}
    \label{tab:Complexity}
    \vspace{-3mm}
\end{table}

\textbf{Ablation study on tuning strategy}. We compare the potential 7 tuning strategies for DMIC in this subsection. As shown in Table~\ref{tab:Tune}, we set the pre-trained model as a performance baseline. The experiments demonstrate that tuning the A2C can obtain the highest performance with only 2.7\% parameters. Specifically, by comparing (b) with (d), we can find tuning the whole model will bring the performance drop since it will destroy prior knowledge existed the well pre-trained feature extractor. From (e), (f), (g), (h), we can find the adapter-style methods are not proper for the RL agent in DMIC. 
  
 \textbf{Comparison with hand-crafted schemes} In this part, we discuss the advantage of the proposed DMIC against hand-crafted schemes. The hand-crafted QP map is generated by utilizing linear, exponential, square, log, and square root functions to establish the relationship between QP value and semantic mask ratio S.
The experimental results are shown in Fig~\ref{fig:RD}, we can find that hand-crafted schemes are far from our proposed DMIC, since our scheme can capture the optimal relationship between QP value and semantic importance adaptively. 

\subsubsection{Complexity Analysis}
In this section, we compare our DMIC with standard software HM 16.24 from the perspective of time complexity. As shown in Table~\ref{tab:Complexity}, our DMIC does not increase any decoding time. For encoding time, our HRL agents only take about 1.52 seconds for the QP decision of the whole video (33 frames) with the size of 512x512. It is efficient and effective to apply our algorithm in task-driven video semantic coding.

\section{Conclusion}
In this paper, we propose the first data-efficient transfer learning-based codec, DMIC, leveraging prior knowledge for few-shot medical diagnosis image compression. To obtain data-efficient transfer learning model, several tuning strategies are ablated based on a pretrained HRLVSC model. Specifically, we focus on tuning merely partial parameters of the full model, which is utilized for diagnosis-oriented bit allocation in the few-shot medical image compression settings. The policy network of DMIC is composed of 
the feature extractor, the fully connected layer, and the reinforcement learning module, which are both ablated or modified. Experiments demonstrate that our DMIC can save 47.594\% BD-Rate compared with the HEVC anchor, and only tuning the A2C can achieve comparable performance with only 2.7\% parameters. In the future work, we will extend our work into more medical tasks, and seek the unified DMIC method for different medical tasks.  


\label{sec: conclusion}
\bibliographystyle{IEEEtran}
\bibliography{reference}
\end{document}